\UseRawInputEncoding
\documentclass[journal=jacsat,manuscript=article]{achemso}

\usepackage{hyperref}

\usepackage[version=3]{mhchem} % Formula subscripts using \ce{}

\author{Sunny Tiwari}
\email{sunny.tiwari@students.iiserpune.ac.in}
\affiliation{Department of Physics, Indian Institute of Science Education and Research, Pune-411008, India}
\author{Utkarsh Khandelwal}
\affiliation{Department of Physics, Indian Institute of Science Education and Research, Pune-411008, India}
\author{Vandana Sharma}
\affiliation{Department of Physics, Indian Institute of Science Education and Research, Pune-411008, India}
\author{G.V. Pavan Kumar}
\email{pavan@iiserpune.ac.in}
\affiliation{Department of Physics, Indian Institute of Science Education and Research, Pune-411008, India}

\title{Single Molecule SERS in a Single Gold Nanoparticle-driven Thermoplasmonic Tweezer}


\begin{document}

%\begin{tocentry}

%\centering
%\includegraphics{Tp.png}
%
%\end{tocentry}

\begin{abstract}
Surface enhanced Raman scattering (SERS) is optically sensitive and chemically specific to detect single molecule spectroscopic signatures. Facilitating this capability in optically-trapped nanoparticles at low laser power remains a significant challenge. In this letter, we show single molecule SERS signatures in reversible assemblies of trapped plasmonic nanoparticles using a single laser excitation (633 nm). Importantly, this trap is facilitated by the thermoplasmonic field of a single gold nanoparticle dropcasted on a glass surface. We employ bi-analyte SERS technique to ascertain the single molecule statistical signatures, and identify the critical parameters of the thermoplasmonic tweezer that provide this sensitivity. Furthermore, we show the utility of this low power ($\approx$0.1mW/$\mu$m$^2$) tweezer platform to trap single gold nanoparticle and transport assembly of nanoparticles. Given that our configuration is based on a dropcasted gold nanoparticle, we envisage its utility to create reconfigurable plasmonic metafluids in physiological and catalytic environments, and can be potentially adapted as an \textit{in-vivo} plasmonic tweezer.
\end{abstract}


Optical confinement and interrogation of single molecules in a fluid remains one of the vital goals of nanoscience and technology\cite{doi:10.1021/acs.nanolett.0c04042,  kulzer2004single, wang2012probing, bishop2018protein, bespalova2019single, franzl2019thermophoretic}. This letter discusses how single molecule surface enhanced Raman scattering (SERS) can be performed in a single gold nanoparticle-driven thermoplasmonic trap of plasmonic particles. \par

Fluorescence and Raman scattering have been extensively used to study molecules in a fluid. Generally, fluorescence is sensitive enough to detect single molecules in a fluid, but Raman scattering remains a weak process due to low optical cross-sections\cite{kulzer2004single}. An important way to overcome this problem is to use SERS\cite{le2008principles, langer2019present, kumar2012plasmonic, lim2010nanogap, doi:10.1021/acscentsci.8b00423}, a spectroscopic method that is sensitive enough to detect single molecules in a fluid phase\cite{le2008principles, le2012single, patra2014plasmofluidic, ahmed2012single, haran2010single, weiss2001time, zrimsek2013single, dieringer2007frequency}. An important prerequisite for single molecule SERS (SM-SERS) detection is the enhanced electromagnetic fields\cite{liebel2020surface, alvarez2011gold, pozzi2015evaluating, kleinman2011single}. Plasmonic structures have been extensively designed and utilized for near-field interaction with the molecules\cite{roy2021controlling, punj2015self,zhang2019unraveling, de2017optimization, dominguez2016adsorption}, and thereby influencing their optical emission not only in terms of their intensity\cite{chikkaraddy2016single, russell2012large, punj2013plasmonic} but also in terms of their directivity\cite{vasista2018differential, doi:10.1021/acs.jpclett.1c01923}.
Optical trapping\cite{jones2015optical, marago2013optical, dienerowitz2008optical, doi:10.1021/jz501409q} and assembly of nanoparticles in a fluid\cite{nan2019silver,han2020phase, gargiulo2017understanding, yan2013optical} are one of the effective ways to create large-scale electromagnetic enhancement with reconfigurable nano-constructs that can be deterministically transported within the fluidic environment\cite{lin2016light, ghosh2018mobile}. 

In this context, performing SERS in an optical or an optothermal trap has emerged as an interesting prospect\cite{lin2016light, fazio2016sers, messina2011manipulation, kang2016surface}, and there have been some attempts towards this end, including our work on single-molecule SERS in an evanescent plasmonic trap\cite{patra2014plasmofluidic}.
Many of the SERS studies in optical traps require multiple lasers operating at relatively high powers\cite{tong2009optical, svedberg2006creating}. For single molecule studies, this may not be a conducive condition. Plasmonic tweezers\cite{kotsifaki2019plasmonic,doi:10.1021/acs.nanolett.0c00300, doi:10.1063/5.0032846, crozier2019quo, shoji2014plasmonic, tsuboi2010optical, masuhara2021optical} have emerged as an alternative to trap nanoscale entities with low power budget\cite{zhang2021plasmonic, li2021algorithm}. In recent times, the thermoplasmonic fields around a metal film\cite{lin2018opto, lin2016light} or other designed nanostructures\cite{liu2018nanoradiator, yoo2018low, sharma2020large} have been utilized to optically trap nano and micro objects at extremely low powers.\par

Although, a large number of studies have shown the capabilities of thermoplasmonic tweezer\cite{wang2011trapping, jones2018photothermal, ndukaife2016long, hong2020stand, braun2015single} to manipulate nanoscale objects, there is still a requirement of tweezer strategies that can be mobile in a fluid and adaptable in biological conditions such as cellular interiors. One of the simplest possible thermoplasmonic nanostructure\cite{baffou_2017, baffou2020applications} that can be introduced in a variety of environment is a chemically prepared gold nanoparticle that can be directly dropcasted from solution phase\cite{baffou_2017, setoura2017stationary, chikazawa2019flow}. Furthermore, it is vital to test the capability of such gold nanoparticles to act like a trapping location for SM-SERS studies. \par

Motivated by these requirements, in this letter we show how SM-SERS can be performed in a thermoplasmonic trap driven by a single, dropcasted gold nanoparticle. This dropcasted nanostructure acts as an anchor trap to create an assembly of plasmonic particles which can be further utilized for SM-SERS studies. By utilizing bi-analyte\cite{le2006proof} SERS technique, we provide experimental proof of SM-SERS signature in such a thermoplasmonic traps. Furthermore, we show how such nanoparticle based thermoplasmonic trap can be utilized for single nanoparticle trapping and optical transport of nanoparticle assemblies, all at a very low laser power density.

\begin{figure}[h!]

\centering
\includegraphics[width=\linewidth]{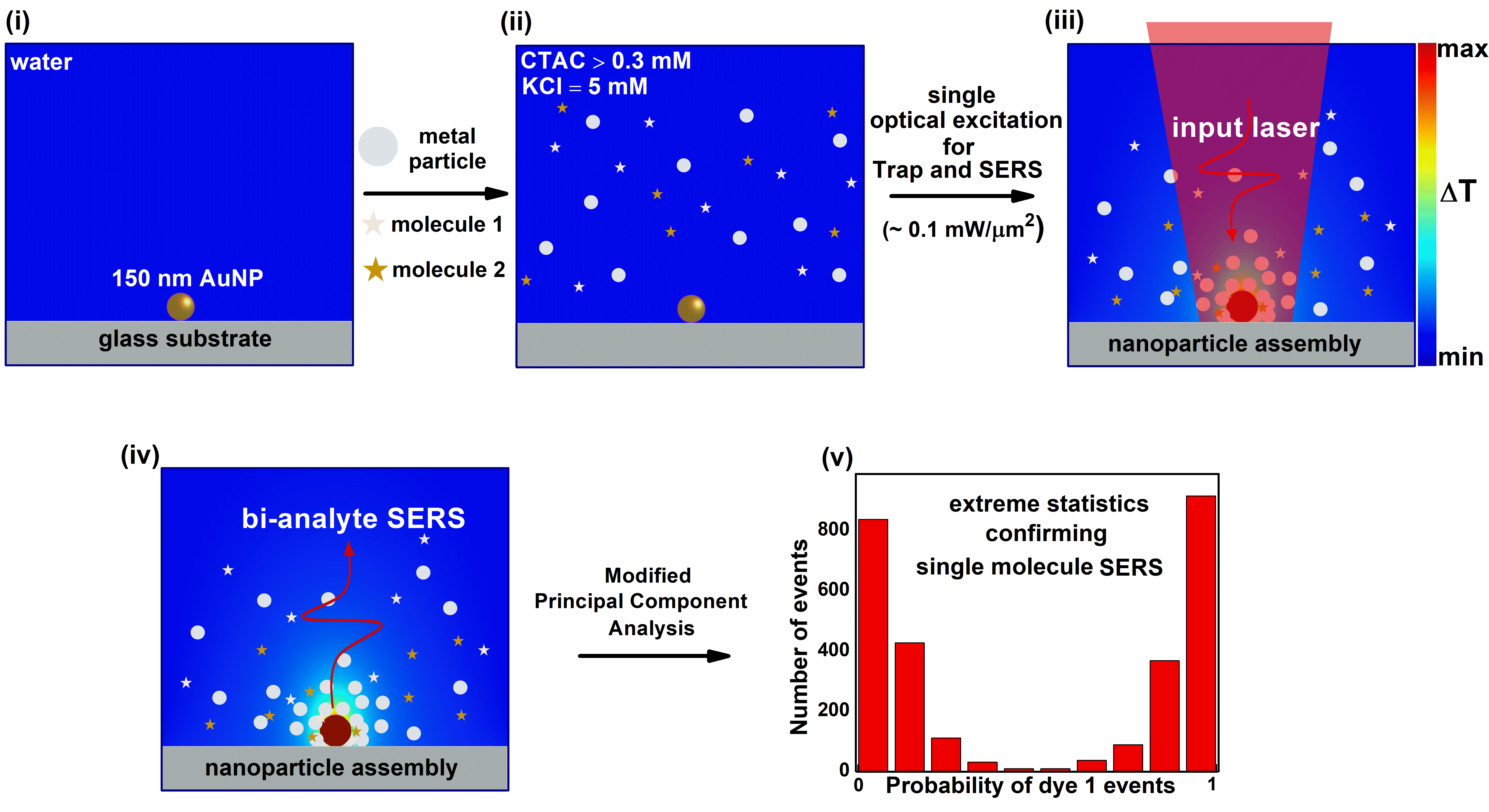}
\caption{Schematic of the experimental configuration to perform SM-SERS in a low power, single nanoparticle driven thermoplasmonic tweezer. (i) A single 150 nm AuNP dropcasted on a glass substrate to be operated as an anchor particle for trapping. (ii) The AuNP was surrounded by a solution containing CTAC and KCl mixed with molecules and plasmonic particles (AuNPs or Lee-Meisel AgNPs). (iii) A low power density laser source ($\approx$0.1mW/$\mu$m$^2$) was used to heat the AuNP for achieving reversible assembly of plasmonic particles. (iv) SERS was performed on the molecules in the assembled plasmonic particles using the same laser source. (v) Modified principal component analysis was performed to statistically confirm SM-SERS signature.}
\end{figure}

A schematic of the experimental configuration is shown in figure 1. A single 150 nm diameter gold nanoparticle (AuNP) was dropcasted on a glass substrate to work as an anchor particle to create an assembly of plasmonic structures. The AuNP was surrounded with an aqueous solution of plasmonic particles, molecules, surfactant cetyltrimethylammonium chloride (CTAC) and potassium chloride (KCl). The plasmonic particle assembly provides the electromagnetic enhancement for the SM-SERS \cite{le2008principles}. The CTAC and the KCl were used as assisting agents for thermoplamsonic trapping and SM-SERS, respectively. We discuss their role later in the manuscript. A single, low power-density laser was used to simultaneously heat the anchor metal nanoparticle\cite{baffou_2017} for thermoplasmonic trapping of plasmonic particles, and as an excitation source for SERS. For SM-SERS validation, we used bi-analyte technique.  Two types of molecules were added in the solution after which modified principal component analysis (MPCA) was performed to statistically confirm the SM-SERS signature\cite{etchegoin2007statistics}. See Supporting information  S1 for detailed schematic of optical microscope, and S2 for discussion on mechanism of the reversible trap.

\begin{figure}[h!]
\centering
  \includegraphics[width=\linewidth]{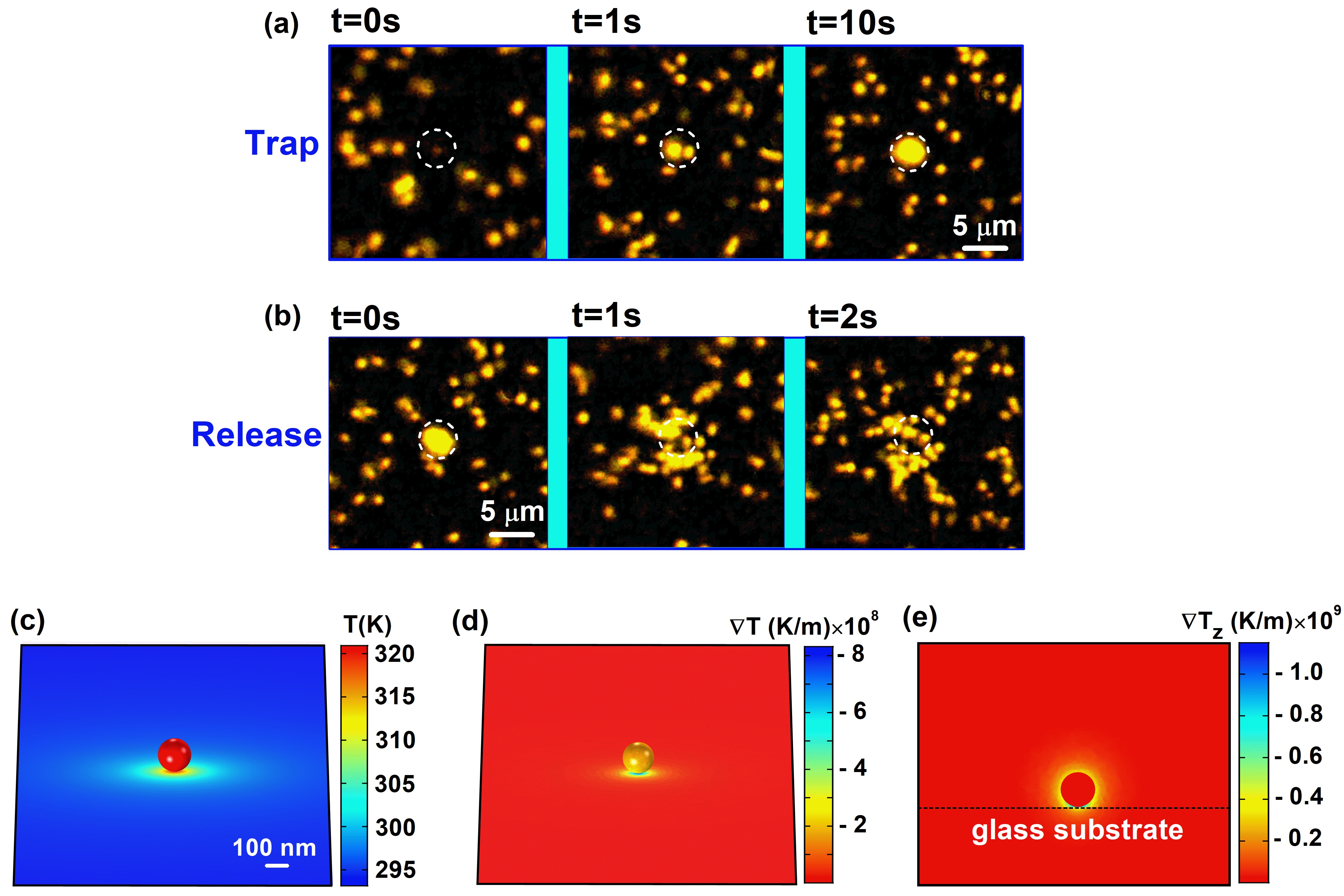}
  \caption{Reversible trapping of plasmonic particles in the low power thermoplasmonic tweezer. Time series dark field images of (a) assembly  of 250 nm AuNPs and (b) disassembly of the same AuNPs when the laser was turned off. The dotted, white circle represents the location of the anchor 150 nm AuNP. Excitation at 633 nm laser with a power density of 0.08 mW/$\mu$m$^2$. (c) Calculated temperature distribution around a 150 nm AuNP placed on a glass substrate and surrounded by water. Excitation at 633 nm laser with a power density of 0.08 mW/$\mu$m$^2$ (d) Calculated temperature gradient $\nabla$T  for the same geometry. (e) Calculated temperature gradient $\nabla$T in the z-plane. See Supporting Movie 1 for trapping and release of 250 nm AuNPs.}
\end{figure}

First, we discuss the experiments on reversible trapping of 250 nm AuNPs in a CTAC solution. Figure 2(a) and (b) show the trapping and release sequence of 250 nm AuNPs.
A 150 nm AuNP was excited by focusing a 0.74 mW laser of wavelength 633 nm at a 3.34 µm spot size by using a 50x, 0.50 NA objective lens. The AuNP was surrounded with an aqueous solution having 12 mM CTAC. As the laser was turned on, the plasmonic particles were seen to be trapped near the anchor AuNP and within 10 s, an assembly of size greater than 5 $\mu$m was achieved. The assembly disappears within a few seconds after turning off the laser. This process can be reproduced over multiple cycles. See Supporting Movie 1 for trapping and release of 250 nm AuNPs.\par

In the absence of the CTAC, no trapping of plasmonic particles was observed at the same input power density\cite{lin2016light}. The particles get scattered away because of the large scattering force when they enter the excitation volume (see Supporting Movie 2). Assembly of nanoparticles of size as large as 400 nm were also created efficiently at low power density, using the single nanoparticle driven thermoplasmonic trapping (see Supporting Movie 3). A slightly more efficient trapping at 532 nm was achieved with a laser source of 532 nm which is nearer to the absorption maxima of the anchor particle. See Supporting Information S3 and S4 for the details on the measurement of scattering cross-section and calculation of absorption cross-section of 150 nm AuNP placed on a glass substrate and surrounded with water, and trapping experiments performed with a 532 nm laser. Note that our thermoplasmonic tweezer is mainly driven by a single anchor AuNP on a glass substrate, in contrast to metal film-based tweezers \cite{lin2018opto}. \par
To further understand the optothermal effects \cite{lin2018opto} in our tweezer, we performed finite element method simulations (COMSOL). The refractive index of gold was taken from ref.\cite{johnson1972optical} (see Supporting Information S5 for details on numerical calculations). Figure 2(c) shows the calculated temperature distribution of a 150 nm AuNP placed on a glass substrate and surrounded with water.  A temperature increment of $\approx$27 K was obtained by exciting the AuNP with a laser of power density 0.08 mW/$\mu$m$^2$. In addition, there was a temperature gradient, both in-plane and out-of-plane of our geometry, because of the heating of AuNP as shown in figure 2(d) and (e), respectively.  This gradient has an important connection to CTAC in the solution.
The plasmonic particles in the fluid and the anchor particles were coated with a positive CTAC double layer because of the addition of CTAC in the solution\cite{gomez2012surfactant}. Since, the concentration of CTAC is greater than critical micelle concentration(0.3 mM), the solution also contains CTAC micelles and negatively charged chlorine ions.
The temperature gradient creates a thermophoretic movement\cite{reichl2014charged} of CTAC double layer coated plasmonic particles, and CTAC micelles and Cl$^-$ ions present in the solution and facilitates the reversible trapping of metallic structure. The particles get trapped near the heated anchor AuNP because of balancing of the thermogradient force between negatively charged Cl$^-$ ions and positively charged CTAC micelles, and the electrostatic repulsive force between the positive coating of CTAC double layer on the anchor particle and CTAC double layer coated metal particles.

\begin{figure}[htbp!]
\centering
\includegraphics[width=\linewidth]{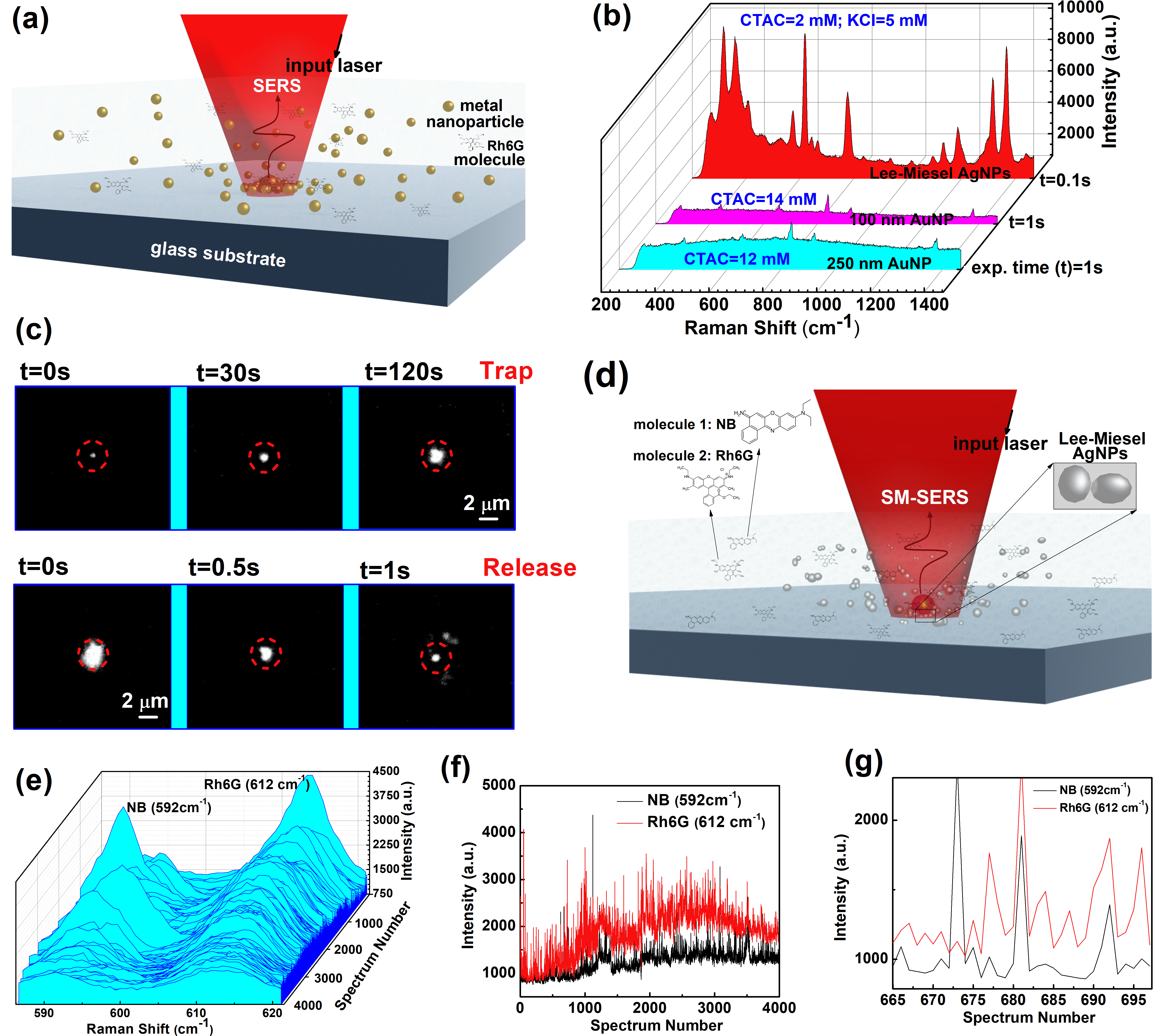}
  \caption{Optimization of experimental configuration for SM-SERS. (a) Schematic of the SERS performed for different plasmonic particles (250 nm and 100 nm AuNPs, and Lee-Meisel AgNPs) with Rh6G molecules. (b) Compassion of the SERS spectra at same input power using 250 nm AuNPs, 100 nm AuNPs and Lee-Meisel AgNPs (with varied concentrations of CTAC and KCl). Time series images of (c) assembly and disassembly of the Lee-Meisel AgNPs. (d) Schematic of SM-SERS using bi-analyte method at 633 nm laser. The trapping configuration included the following: 150 nm anchor AuNP, Lee-Meisel AgNPs in solution, NB and Rh6G molecules in the presence of CTAC (2 mM)  and KCl (5 mM). (e) 4000 bi-analyte SERS spectra recorded from a solution containing 1 nM of both NB and Rh6G molecules. The integration time for single spectrum was 0.3 s and the dwell time between two consecutive spectra was 1 s. (f) SERS intensity variation of 592 cm$^{-1}$ and 612 cm$^{-1}$ peaks of NB and Rh6G molecules, respectively, as a function of spectrum number. (g) Same as the (f), but for a smaller window of spectral numbers indicating higher resolution. See Supporting Movie 4 and 5 for trapping and release of Lee-Meisel AgNPs respectively.}
\end{figure}

Having shown the formation of large scale assembly, next we show how the assembly can be utilized for performing SERS of rhodamine 6g (Rh6G) molecules and compare its performance for various plasmonic particles. A schematic of the experimental design is shown in figure 3a. A reversible assembly of plasmonic particles (250 or 100 nm AuNPs or Lee-Meisel AgNPs)  was achieved by heating a 150 nm AuNP. The anchor AuNP was illuminated with a water immersion 1.2 NA, 60x objective lens by focusing a 0.80 mW, 633 nm laser to a spot size of 2.70 µm. (For this optical excitation, the temperature distribution, and gradients are shown in Supporting Information S6 )
After an assembly formation of about the spot size, SERS spectroscopy of Rh6G molecules was performed in the backscattering configuration using the same excitation laser used for the trapping purpose. Figure 3b compares the SERS spectrum collected from 10$^{-6}$M Rh6G molecules in the assembly of 250 nm AuNPs, 100 nm AuNPs and Lee-Meisel AgNPs. 
As can be seen from the SERS spectra figure 3b, the SERS sensitivity was superior for Lee-Meisel AgNPs in the presence of CTAC and KCl. From the past work, it is well known that Lee-Meisel AgNPs are excellent substrates for achieving large SERS enhancement\cite{lee1982adsorption, le2008principles}, when a small amount of KCl is added to the solution \cite{meyer2006self, otto2003chloride, Grochala1998AnioninducedCE}. Therefore, prior to adding CTAC in the solution containing Lee-Meisel AgNPs and molecules, KCl was added in the solution. The addition of KCl initiates dimer formation of AgNPs which creates hotspots for SERS enhancement \cite{meyer2006self}. A low concentration of CTAC (2 mM) was added after formation of dimers, which leaves enough space for the molecules to access the hotspots. 
Figure 3c shows the trap and release of large scale assembly of Lee-Meisel AgNPs using nanoparticle driven thermoplasmonic trap. An assembly of few microns was formed within tens of seconds after the anchor particle was illuminated with the laser source. The assembly disappears very fast after the laser is turned off. See Supporting Movie 4 and 5 for trapping and releasing of Lee-Meisel AgNPs respectively), and SI 7 for details on preparation of Lee-Meisel AgNPs.

To push the limit of single nanoparticle driven thermoplasmonic trap, we performed SM-SERS spectroscopy on the assembly of trapped Lee-Meisel AgNPs using bi-analyte technique, whose details can be found elsewhere \cite{le2006proof, patra2013single, patra2014plasmofluidic}. Figure 3(d) shows the schematic of the experimental configuration to perform bi-analyte SERS experiments using two types of molecules, which compete for the same hotspot\cite{le2006proof}. The solution contains two types of molecules, Nile Blue (NB) (1 nM) and Rh6G (1 nM) mixed with Lee-Meisel AgNPs. As discussed earlier, the illuminated anchor AgNPs drives the assembly and SERS was performed from the assembly in the presence of CTAC and KCl (concentration at 2 mM and 5 mM respectively). Various other control experiments were performed (See Supporting Information S8 and S9 for bi-analyte SERS experiments performed without adding KCl and on a location away from assembly respectively. Also see Supporting Information S10 for SERS experiments performed using a 532 nm laser) \par
Next, we discuss the obtained SERS spectra. A waterfall plot of 4000 spectra containing closely spaced NB SERS peak 592 cm$^{-1}$ and Rh6G SERS peak (612 cm$^{-1}$) is shown in figure 3(e). The exposure time for each spectrum was 0.3s and dwell time between two consecutive spectra was 1 s. The variation in the peak intensity can be seen clearly. Figure 3(f) shows the occurrence and variation in the intensity of both 592 cm$^{-1}$ peak of NB and 612cm$^{-1}$ peak of Rh6G molecules as a function of spectrum number. The zoomed-in portion of spectra from 665 to 697 (figure 3(g)) shows the individual events of both molecules and a mixed event. At an event number 673 only an intense NB peak of 592 cm$^{-1}$ was observed, whereas at event number 681 both peaks (592 cm$^{-1}$ of NB and 612 cm$^{-1}$ of Rh6G molecule) was seen, and at 696, only an intense peak of 612 cm$^{-1}$ of Rh6G peak was visualized. These fluctuating individual events suggest that the single nanoparticle driven thermoplasmonic trap can be used to perform SM-SERS.

\begin{figure}[H]
\centering
  \includegraphics[width=\linewidth]{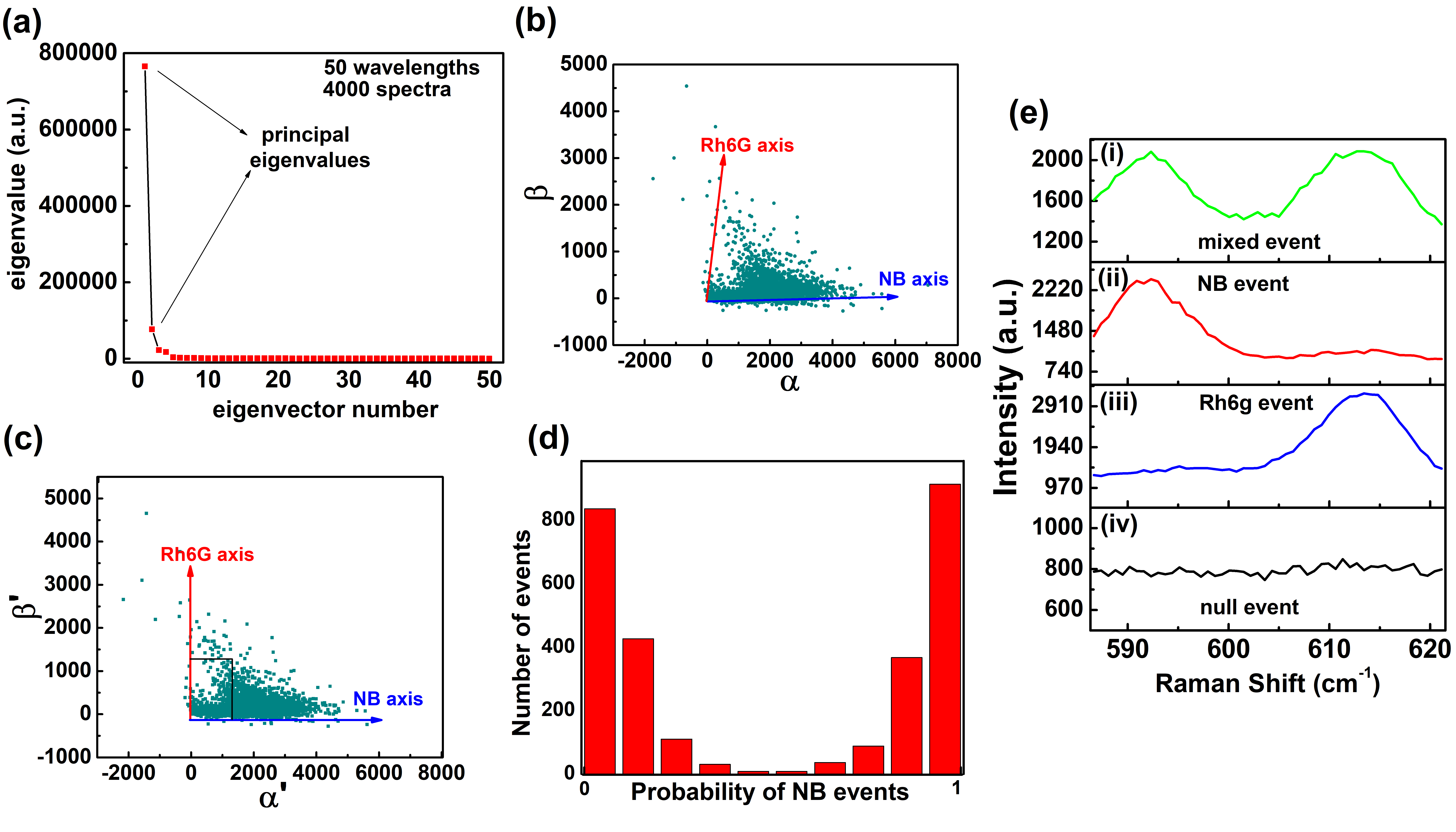}
  \caption{ Modified principal component analysis of bi-analyte SM-SERS. (a) Eigenvalues obtained from PCA of 4000 spectra consisting of 50 wavelengths containing the 592cm$^{-1}$ and 612 cm$^{-1}$ peaks of NB and Rh6G, respectively. (b) Distribution of the covariant matrix coefficients ($\alpha$ and $\beta$) obtained after PCA showing two Rh6G and NB axes. (c) Distribution of the transformed matrix coefficients $\alpha'$ and $\beta'$ showing two orthogonal Rh6G and NB axes. (d) Probability of NB events plotted using (c) after removing the mixed dye events shown in a black region in (c). (e) Four possible events: (i) Mixed event obtained near the center of the histogram (d). (ii)  NB event obtained near the region of probability of NB event=1 (iii) Rh6G event obtained near the region of probability of NB event=0. (iv) No event obtained when probabilities of both NB event and Rh6G event are 0.}
\end{figure}

To  statistically confirm the single molecule SERS signature, we performed modified principal component analysis (MPCA)\cite{etchegoin2007statistics, patra2013single, patra2014plasmofluidic} on 4000 spectra collected (figure 3(e)) from the solution containing both NB and Rh6G molecules. Figure 4a shows the plot of calculated eigenvalues and eigenvectors, where the first two elements are the principal eigenvalues. Figure 4b shows the scatterplot of two covariant matrix coefficients $\alpha$ and $\beta$, with two axes shown as NB and Rh6G. Each data point in the scatterer plot represents an event. The re-oriented scatterer plot ($\alpha'$ and $\beta'$) in figure 4c shows two orthogonal axes Rh6G and NB axes, where events on both axes represent pure events of SERS from only one type of molecule. Events in between the two axes represent mixed events which are signals from both the types of molecules. To calculate the frequency of single molecule events, a histogram was plotted in figure 4d where the frequency of only NB molecule was plotted after removing noisy mixed events as shown by a black box in figure 4c. The extreme left part of the histogram represents the events where SERS from only Rh6G molecules was recorded. The spectrum in figure 4 (e)(iii) shows such a type of an intense pure event from only Rh6G molecules. The center part of the histogram shows the mixed events from both types of molecules and one such event is shown in figure 4(e)(i). The extreme right part of the histogram displays the probability of only NB molecule events and one such event is shown in figure 4(e)(ii). Along with these three events, a relatively less probable null event was also observed, which is shown in figure 4(e)(iv). For further details on the MPCA analysis, see Supporting Information S11.

\begin{figure}[H]
\centering
  \includegraphics[width=\linewidth]{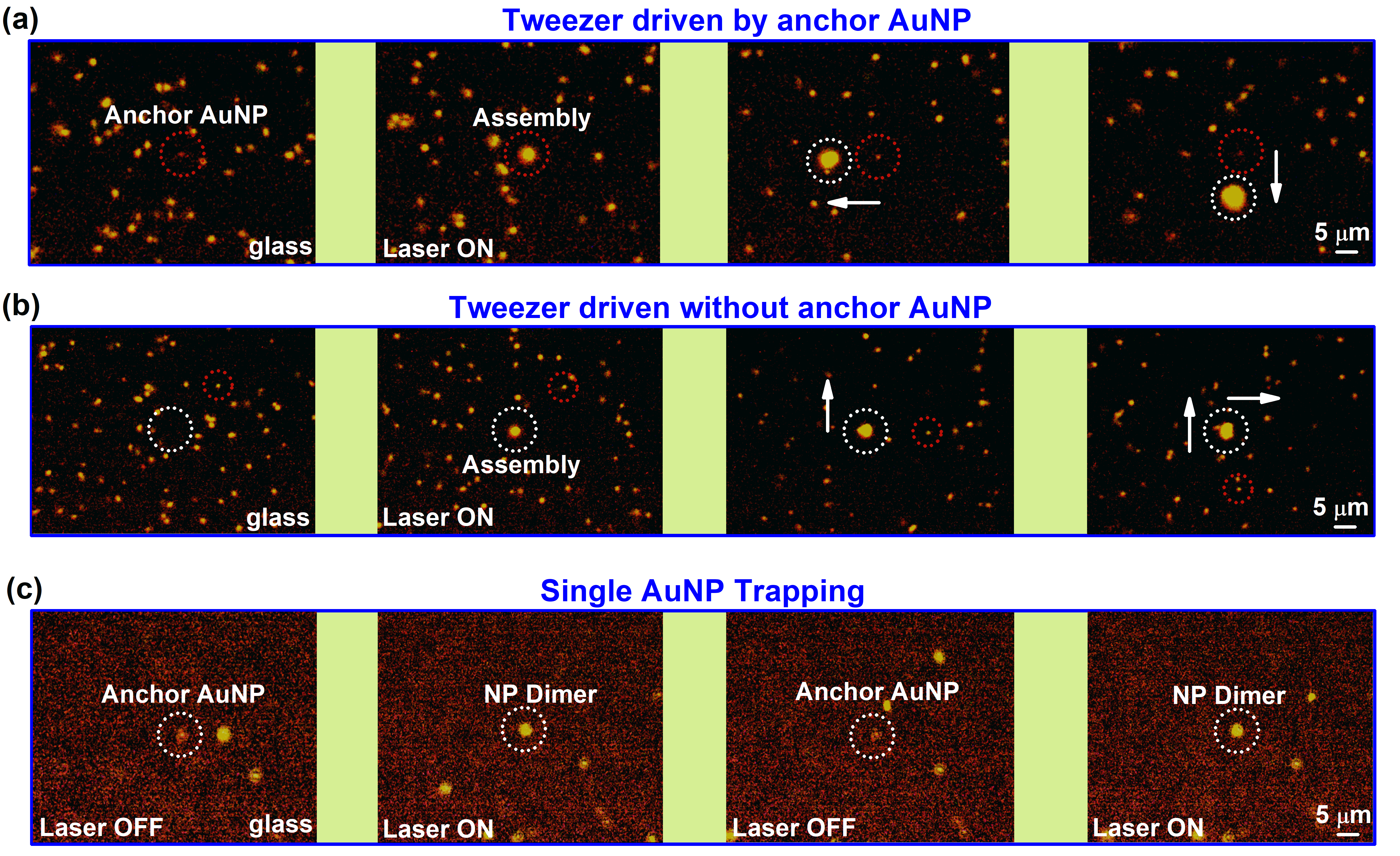}
  \caption{Versatility of nanoparticle driven thermoplasmonic tweezer. (a) Tweezer driven by anchor AuNP. Time series dark field images of assembly and its translation along two axes. (b) Tweezer driven without anchor AuNP. Time series dark field images of assembly and its translation along two axes. Note that this assembly process is stochastic and takes longer duration compared to the previous case. (c) Single particle trapping and release. A reversible dimer formation was observed. Excitation wavelength: 633 nm at a power density of 0.08 mW/$\mu$m$^2$, anchor AuNP: 150 nm, trapped plasmonic particles: 250 nm AuNPs, for all the observations. See Supporting Movie 6,7 and 8 related to the above data.}
  
\end{figure}

Can our tweezer platform assemble and transport plasmonic structures? Can such a platform trap single nanoparticles at the low laser power densities? In order to address these questions, further experiments were performed as shown in figure 5. First, we trapped and transported an assembly of 250 nm AuNPs along two different axes, as shown in figure 5a. We performed this transport by holding the trap at its position and moving the motorized-stage of the optical microscope. An interesting aspect of this transport is that once assembled, the plasmonic particles do not need the assistance of the anchor nanoparticles for further trapping. This may be due to temperature increment and gradient facilitated by the assembly and hence acting as a ‘self-sufficient’  trap. See Supporting Movie 6 showing the working of tweezer driven by anchor AuNP.
Secondly, we show that the assembly can also be created without the assistance of anchor NP (figure 5b).  Although we could trap and transport the assembly of plasmonic particles, we found the trapping process to be stochastic and can take much longer times when compared to anchor AuNP based tweezers (figure 5a). The figure 5(b) shows the formation of the assembly when the laser was illuminated on a glass substrate. Initially, a single particle comes under the excitation volume and gets pushed towards the glass substrate because of scattering force, and it stays there for finite duration. Within that time, the temperature of the particle increases, which leads to the formation of a thermal gradient, and hence a trap. See Supporting Movie 7 showing the working of tweezers driven without anchor nanoparticles.

Finally, we show how a single 250 nm plasmonic particle can also be trapped using a single, anchor AuNP, thus forming a dynamic, plasmonic dimer. Figure 5(c) shows the reversible trapping and release of a single 250 nm AuNP. Turning off the laser results in the release of the trapped particle from the anchor location. Turning on the laser again, further traps the nanoparticle thus confirming the reversibility. See Supporting Movie 8 showing single nanoparticle trapping.

To conclude, we have shown how a single dropcasted gold nanoparticle on a glass substrate can be utilized as a thermoplasmonic tweezer to reversibly assemble plasmonic nanoparticles for SM-SERS experiments. By employing bi-analyte SERS technique we have revealed the single-molecule spectral dynamics and its extreme statistical signatures in the assembly of nanoparticles. We have emphasized the relevance of surfactant and salt solution in creating a conducive environment for trapping and SM-SERS at low laser powers ($\approx$0.1mW/$\mu$m$^2$) . Furthermore, we have revealed the capability of the tweezer to not only trap single plasmonic nanoparticles, but also to transport plasmonic nanoparticle assembly to a desired location. Such an assembled plasmonic entity can be potentially harnessed as components of plasmonic meta-fluid. Given that our tweezer platform is based on a colloidal gold nanoparticle, it can be chemically functionalized for targeted trapping of nano-objects including bio-molecules. We envisage its utility as a flexible trapping and transport platform \textit{in-vivo}, and anticipate further exploration on colloidal plasmonic tweezers sensitive to single-molecule reactions in physiological conditions.

\begin{acknowledgement}

Authors thank Naveen Nishad and Angira Rastogi (IISER Pune) for helping with the MPCA code, and Debashree Roy, Diptabrata Paul, and Shailendra K Chaubey (IISER Pune) for fruitful discussion. Authors also thank Dr. Rohit Chikkaraddy (University of Cambridge) and Dr. Partha Pratim Patra (Dispelix Oy) for a discussion on MPCA, and Gokul MA and Pranab Dutta for helping in experiments. 
This work was partially funded by Air Force Research Laboratory grant (FA2386-18-1-4118 R{\&} D18IOA118), DST Energy Science grant (SR/NM/TP-13/2016), and Swarnajayanti fellowship grant (DST/SJF/PSA02/2017-18) to G V PK.

\end{acknowledgement}

\begin{suppinfo}

The Supporting Information is available free of charge.

Experimental setup, mechanism of trap, scattering and absorption cross-sections, details on FEM simulations, details on MPCA, controlled experiments on bi-analyte SERS with 633 and 532 nm laser. (pdf)

The following videos can be found here 

\url{https://www.youtube.com/playlist?list=PLVIRTkGrtbrvs7BaNsaH6tjPpzLUizyMI}

Reversible trapping of 250 nm AuNPs (AVI), 
Effect of CTAC on trapping (AVI), 
Reversible trapping of 250 nm AuNPs (AVI), 
Trapping of Lee-Meisel AgNPs (AVI), 
Release of Lee-Meisel AgNPs assembly (AVI), 
Working of tweezer driven by anchor AuNP (AVI), 
Working of tweezer driven without anchor AuNP (AVI), 
Single AuNP trapping (AVI), 
Reversible trapping of 250 nm AuNPs (AVI), 
Unprocessed movie of reversible trapping of Lee-Meisel AgNPs (AVI).
\end{suppinfo}

\bibliography{achemso-demo}

\pagebreak

%\onecolumngrid
\begin{center}
  \textbf{\large Supplementary Material\\Single Molecule SERS in a Single Gold Nanoparticle-driven Thermoplasmonic Tweezer}\\[.2cm]
  Sunny Tiwari,$^{1, *}$ Utkarsh Khandelwal,$^{1}$ Vandana Sharma$^{1}$ and G. V. Pavan Kumar$^{1, *}$\\[.1cm]
  {\itshape ${}^1$Department of Physics, Indian Institute of Science Education and Research, Pune-411008, India}\\
  ${}^*$Email: sunny.tiwari@students.iiserpune.ac.in, pavan@iiserpune.ac.in\\
%(Dated: \today)\\[1cm]
\end{center}
%\twocolumngrid

\setcounter{equation}{0}
\setcounter{figure}{0}
\setcounter{table}{0}
\renewcommand{\theequation}{S\arabic{equation}}
\renewcommand{\thefigure}{S\arabic{figure}}
\renewcommand{\bibnumfmt}[1]{[S#1]}
\renewcommand{\citenumfont}[1]{S#1}

\pagebreak
\tableofcontents

\appendix
\renewcommand{\thesection}{S\arabic{section}}
\section{}
\renewcommand{\thefigure}{S\arabic{figure}}

\subsection{S1: Thermoplasmonic tweezer coupled with SERS setup}
\subsection{S2: Detailed mechanism of reversible trapping}
\subsection{S3: Scattering and absorption cross-sections of nanoparticle on glass substrate and surrounded by water}
\subsection{S4: Trapping experiments performed with a 532 nm laser}
\subsection{S5: Details on FEM based simulations}
\subsection{S6: Calculated temperature and temperature gradient using a 1.2 NA water immersion lens}
\subsection{S7: Details on preparation of Lee-Meisel silver nanoparticles}
\subsection{S8: Bi-analyte SERS without adding KCl in the solution}
\subsection{S9: Bi-analyte SERS away from the assembly}
\subsection{S10: Bi-analyte SERS using 532 nm laser}
\subsection{S11: Details on modified principal component analysis}
\subsection{S12: Details on Supporting Movies}

\pagebreak

\section{S1: Thermoplasmonic tweezer coupled with SERS setup}

\begin{figure}[H]
\centering
  \includegraphics[width=\linewidth]{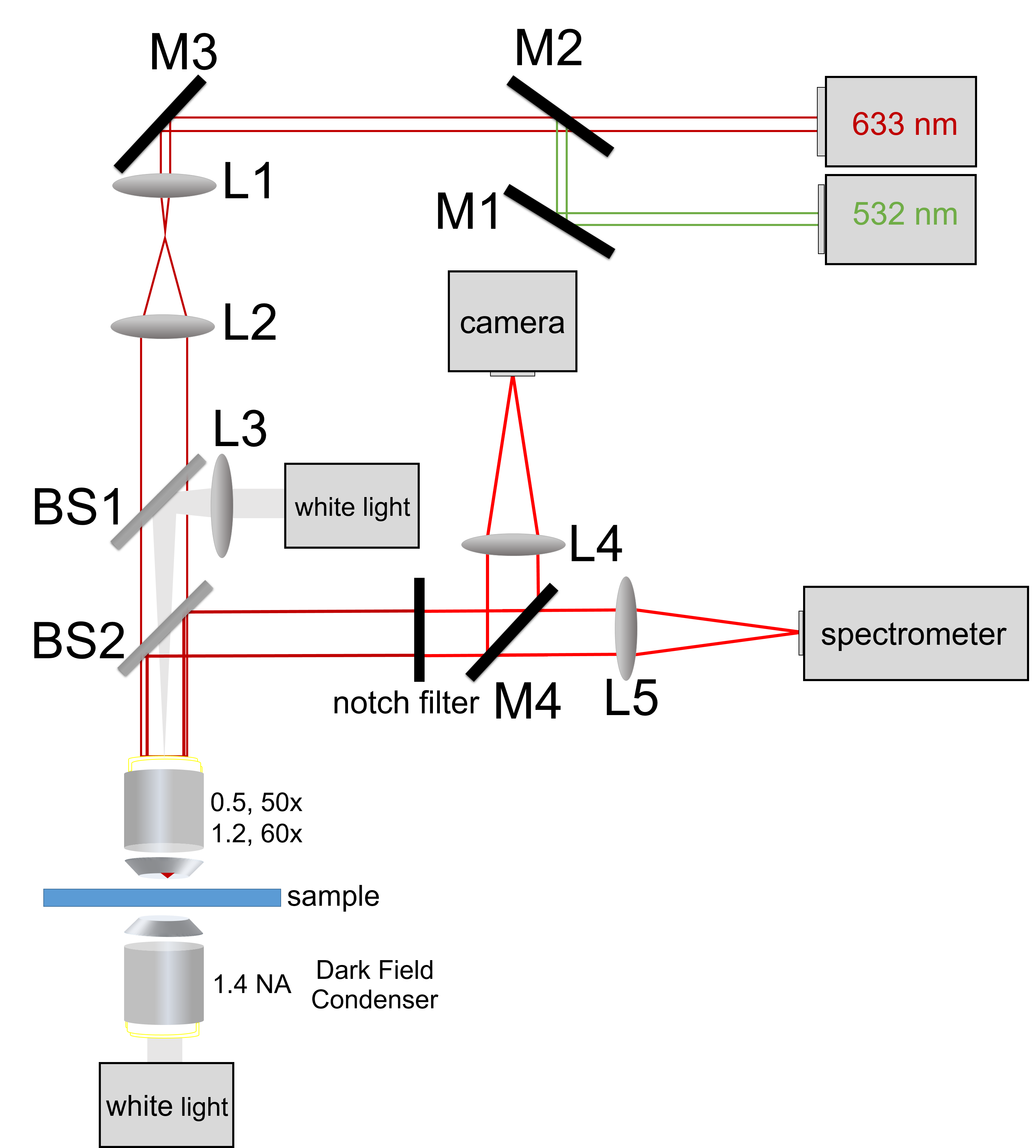}
  \caption{Schematic representation of the setup.}
  
\end{figure}

The sample was illuminated using a 50x, 0.50 NA air objective lens for trapping and 60x, 1.2 NA water immersion objective lens for simultaneous trapping and SERS. The backscattered light was collected using the same lenses.  The laser light was expanded using a set of two lenses L1 and L2. M1 and M3 are mirrors and M2 is a flip mirror to choose either of the lasers. BS1 and BS2 are beam splitters to simultaneously excite the sample with laser and its visualization using white light for bright field imaging. Lens L3 is used to loosely focus white light on the sample plane. Notch filter is used to reject the elastically scattered light for SERS spectroscopy and dark field and bright field imaging. Lenses L4 and L5 are used to focus the emission onto the spectrometer or CCD. M4 is a flip mirror to route light for the imaging or the SERS. For dark field imaging the sample was excited with a dark field condenser from the bottom and the scattered light was collected from the top objective lens. A spacer of height 120 $\mu$m was used while performing trapping experiments and no spacer was used while performing simultaneous trapping and SERS.

\section{S2: Detailed mechanism of reversible trapping}

\begin{figure}[H]
\centering
  \includegraphics[width=\linewidth]{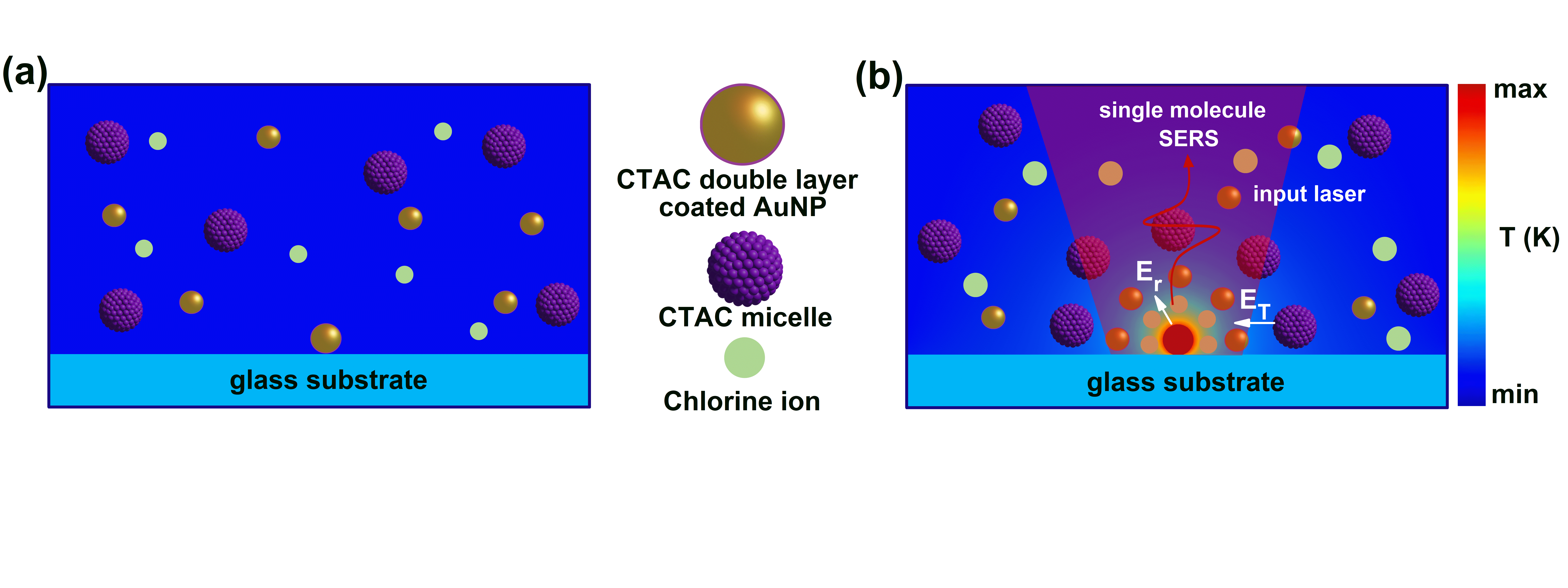}
  \caption{Detailed mechanism of the reversible trapping using single nanoparticle driven thermoplasmonic trap.}
  
\end{figure}

Figure S2 shows the mechanism of the reversible trapping. 150 nm gold nanoparticles in ethanol solution were dropcasted on the glass substrate and were left to dry. The AuNPs were surrounded with an aqueous solution containing plasmonic particles, and CTAC molecules at a concentration greater than critical micelle concentration. The plasmonic particles suspended in the solution and the anchor metal particle get coated with a CTAC double layer which is positive in charge\cite{gomez2012surfactant}. The solution also contains negatively charged chlorine ions. A distribution of the ions is shown in figure S2(a).
As the anchor particle is illuminated with the laser source, the temperature of the particle increases. This also creates a temperature gradient near the anchor metal particle. This temperature gradient leads to the thermophoretic motion of the charged ions in the solution\cite{reichl2014charged}. This causes a redistribution of Cl$^-$ ions and CTAC micelles, and CTAC double layer coated plasmonic particles. The thermoelectric force E$_T$ between the Cl$^-$ ions present near the anchor particle and the positively charged CTAC micelles forces the plasmonic particles to move towards the anchor particle. An interaction between the CTAC coated anchor particle and CTAC micelle creates a repulsive force E$_r$ which balances the E$_T$ and the plasmonic particles get trapped near the anchor particle. After the laser is turned off the repulsive force in addition with the Brownian motion of the plasmonic particles causes the particles to move away from the anchor particle.

\section{S3: Scattering and absorption cross-sections of nanoparticle on glass substrate and surrounded by water}

\begin{figure}[H]
\centering
  \includegraphics[width=\linewidth]{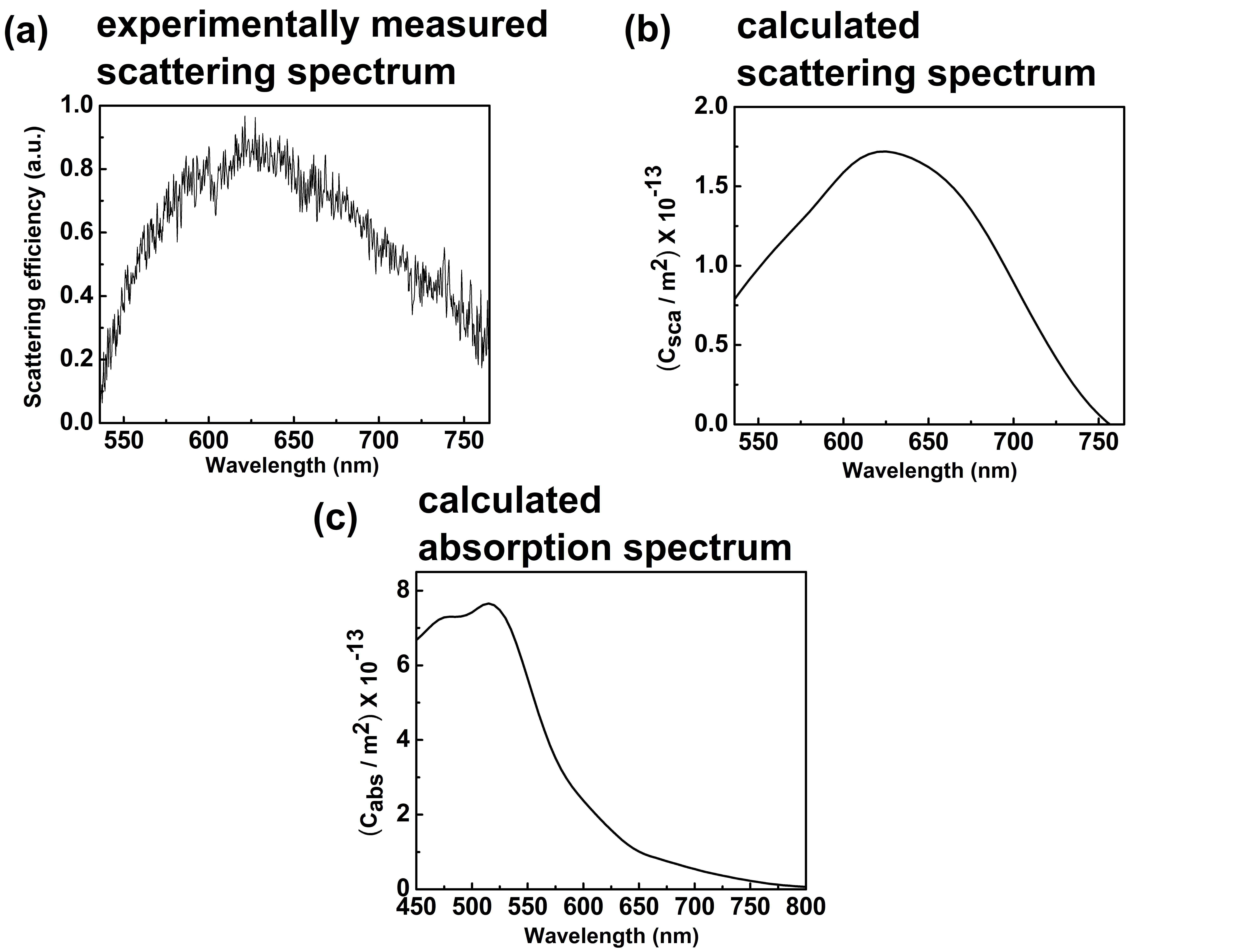}
  \caption{Scattering and absorption cross-sections of a 150 nm AuNP placed on a glass substrate and surrounded by water. (a) Experimentally measured scattering spectrum of the AuNP in water and placed on a glass substrate. (b) Calculated scattering spectrum of a 150 nm AuNP in a material of refractive index 1.365. (c) Calculated absorption of a 150 nm AuNP in a material of refractive index 1.365.}
  
\end{figure}

Figure S3 shows the scattering and absorption cross-sections of a 150 nm diameter nanoparticle. The scattering spectrum was experimentally measured using slant angle excitation and the scattered light was collected using a low numerical aperture objective lens. 
To experimentally calculate the absorption cross-section, the peak of the calculated scattering spectrum was recorded by tuning the refractive of the surrounding medium in which the gold particle was suspended. Refractive index of the medium was chosen by taking the effective refractive index of the medium as the particle was placed on a glass substrate and was surrounded by water\cite{setoura2017stationary}. The peak position of experimentally measured and numerically calculated scattering spectrum matches at an effective refractive index of 1.365. Finally, at a refractive index of 1.365 of the medium the absorption cross-section was calculated and is shown in figure S3 (c). 

\section{S4: Trapping experiments performed with a 532 nm laser} 

\subsection{S4. (i) Calculated temperature and temperature gradient upon illuminating the nanoparticle}

\begin{figure}[H]
\centering
  \includegraphics[width=\linewidth]{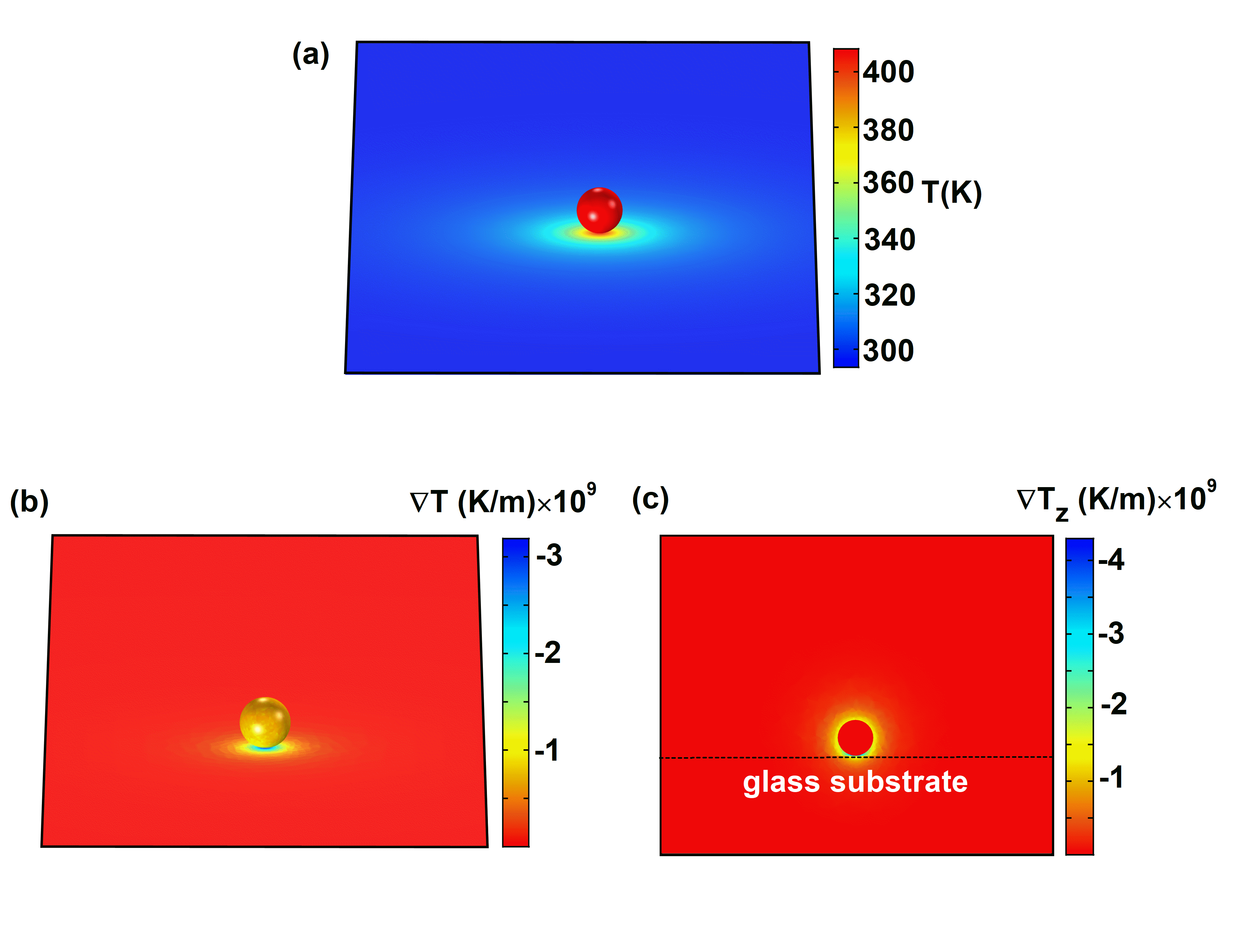}
  \caption{Calculated temperature and temperature gradient upon illumination of a single nanoparticle with a 532 nm laser at a power density of 0.07 mW/$\mu$m$^2$. (a) Temperature increment of the nanoparticle. (b) Calculated temperature gradient ($\nabla$T) in the x-y plane and near the surface of the nanoparticle. (c) Calculated temperature gradient ($\nabla$T ) in the z plane. }
  
\end{figure}

\subsection{S4. (ii) Trapping of 250 nm AuNPs using 532 nm laser}

Movie S9 shows the reversible trapping of 250 nm AuNPs using the single nanoparticle driven thermoplasmonic trap. A single 150 nm diameter was placed on a glass substrate and was surrounded by water, added with CTAC molecules. The solution was placed inside a chamber of height 120 $\mu$m. The final concentration of CTAC in the solution was 15 mM. The particle was illuminated by focusing a 0.55 mW laser of wavelength 532 nm at a 3.17 $\mu$M spot size by using a 50x, 0.50 NA objective lens. 

\subsection{S4. (iii) Effect of CTAC concentration and input power density on the trapping}

\begin{figure}[H]
\centering
  \includegraphics[width=\linewidth]{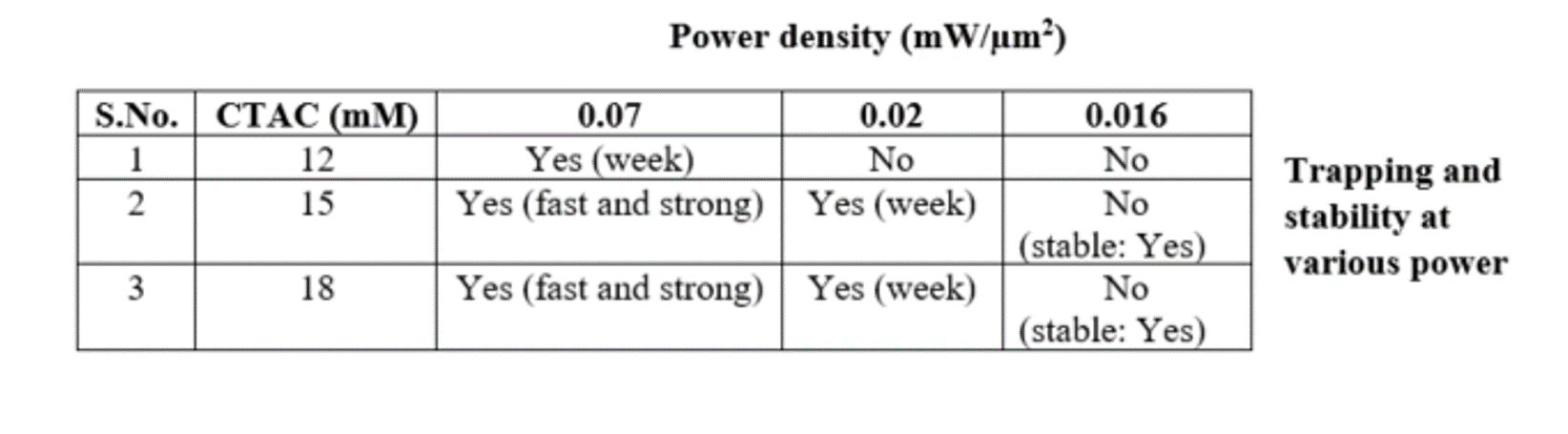}
  \caption{Effect of input power density and CTAC concentration on trapping.}
  
\end{figure}

For a constant CTAC concentration of 12 mM, a week trapping was observed at an input power of 0.07 mW$\mu$m$^2$ but after decreasing the power density to 0.02 mW/$\mu$m$^2$ no trapping was observed.
After increasing the CTAC concentration to 15 mM a fast and strong trapping is observed at an input power density of 0.07 mW/$\mu$m$^2$. A week trapping was also observed at an input power density of 0.02 mW/$\mu$m$^2$. After forming the assembly at a power of 0.07 mW/$\mu$m$^2$, the assembly was stable by reducing the power density to 0.02 mW/$\mu$m$^2$ and also to 0.016 mW/$\mu$m$^2$. A further decrease in the power leads to disassembly. 
After further increasing the CTAC concentration, no observable changes in the trapping was observed.

\section{S5: Details on FEM based simulations}

The heat power density $q(r, t)$at any time $t$  and at any location $r$ inside an illuminated structure is given by\cite{baffou_2017}

\begin{equation}
q(\mathbf{r}, t)=\mathbf{J}(\mathbf{r}, t) \cdot \mathbf{E}(\mathbf{r}, t)
\end{equation}
	  
where $J$ is the electronic current density and $E$ is the electric field inside the nanostructure.

Taking the time average, and denoting the frequency of the light source illuminating the structure as $\omega$, the heat power density becomes,

\begin{equation}
q(\mathbf{r})=\left(\frac{\omega}{2}\right) \varepsilon_{0} \operatorname{Im}(\varepsilon)|\mathbf{E}(\mathbf{r})|^{2}
\end{equation}

Calculation for electric field intensity inside the metal structures and thus temperature increment and temperature gradient upon illumination were performed using finite element method based study using COMSOL Multiphysics as solver.

Electromagnetic power loss density option was selected to account for the heating source. Boundary conditions were defined using the equation 

\begin{equation}
q_{0}=h\left(T_{e x t}-T\right)
\end{equation} 

with $T_{e x t}$ = 293.15 K and $h$ is the heat transfer coefficient, with external natural convection option for a spherical geometry of diameter 150 nm. 
The particle was excited with a focused Gaussian beam with values of power and beam waist size were put in, by measuring the experimental power at the sample plane and beam spot size. The refractive index of the metal particle was taken from ref\cite{johnson1972optical}.

\section{S6: Calculated temperature and temperature gradient using a 1.2 NA water immersion lens}

\begin{figure}[H]
\centering
  \includegraphics[width=\linewidth]{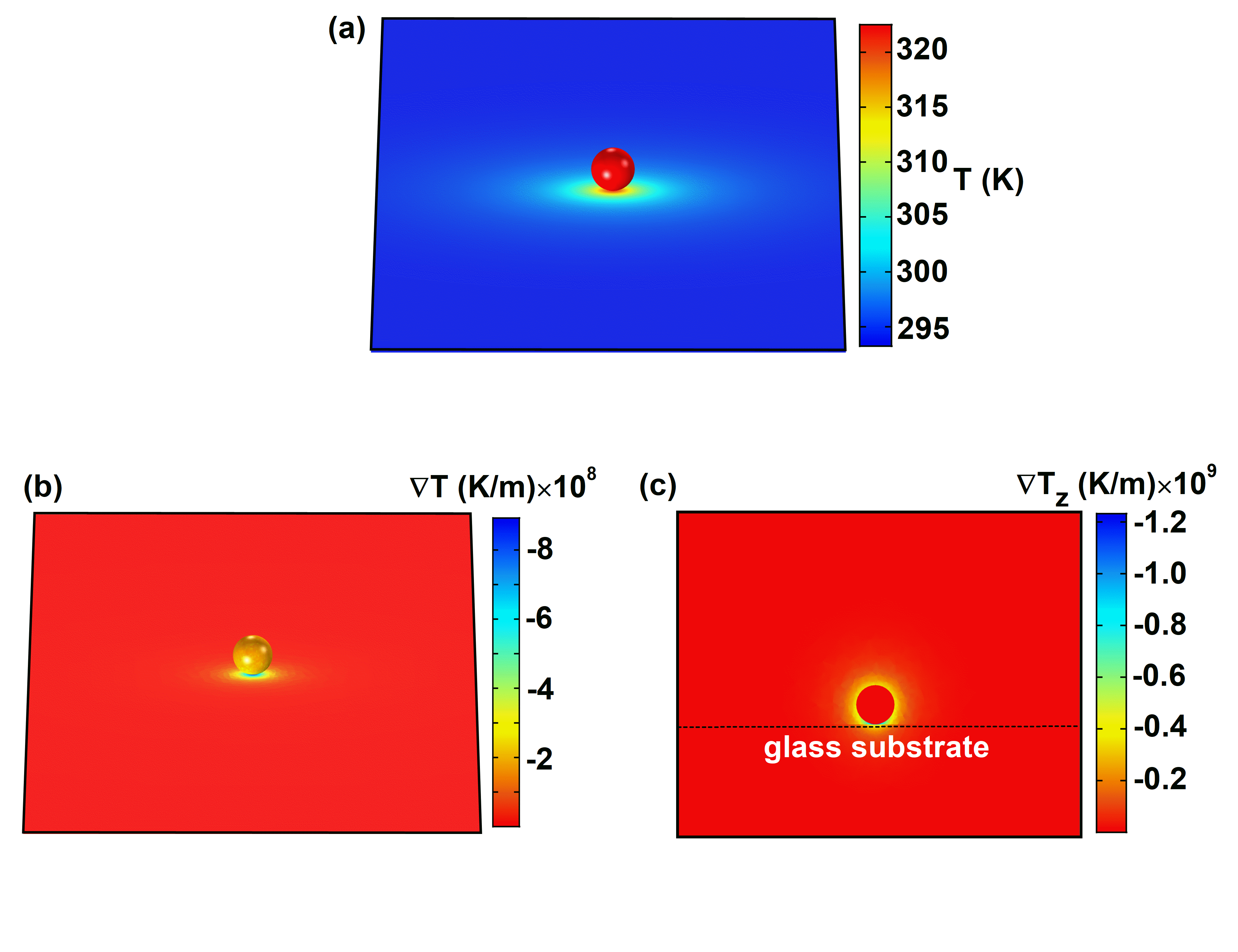}
  \caption{Calculated temperature and temperature gradient upon illumination of a single nanoparticle with a 633 nm wavelength laser at a power density of 0.14 mW/$\mu$m$^2$. (a) Temperature increment of the nanoparticle. (b) Calculated temperature gradient $\nabla$T in the x-y plane and near the surface of the nanoparticle. (c) Calculated temperature gradient $\nabla$T in the z plane.}
  
\end{figure}

\section{S7: Details on preparation of Lee-Meisel silver nanoparticles}

\begin{figure}[H]
\centering
  \includegraphics[width=\linewidth]{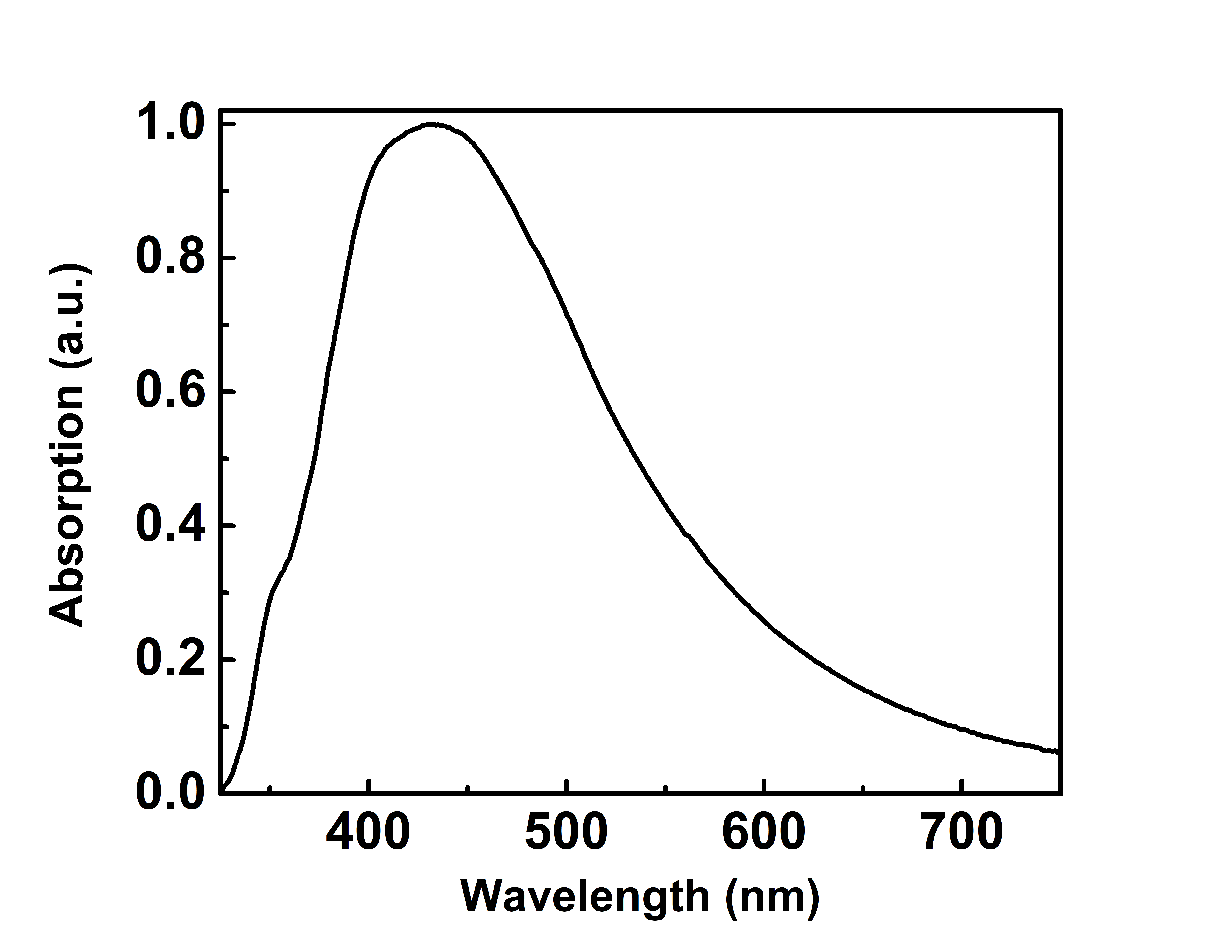}
  \caption{UV-Vis absorption spectra of Lee-Meisel silver nanoparticles.}
  
\end{figure}

180 mg of AgNO$_3$ was dissolved in 1L of mQ water to prepare a 1.04 mM solution. A 10 ml solution of 1$\%$ of Na$_3$C$_6$H$_5$O$_7$ was added while boiling, which serves both as reducing and stabilizing agent. The solution was boiled for one hour and then brought to the room temperature. A gray color solution was obtained which confirms the preparation of silver nanoparticles. The UV-Vis spectrum of the solution is shown in figure S7 and the peak position matches well the reported values\cite{le2008principles}. A detailed synthesis process can be found in ref\cite{lee1982adsorption}.

\section{S8: Bi-analyte SERS without adding KCl in the solution}

\begin{figure}[H]
\centering
  \includegraphics[width=\linewidth]{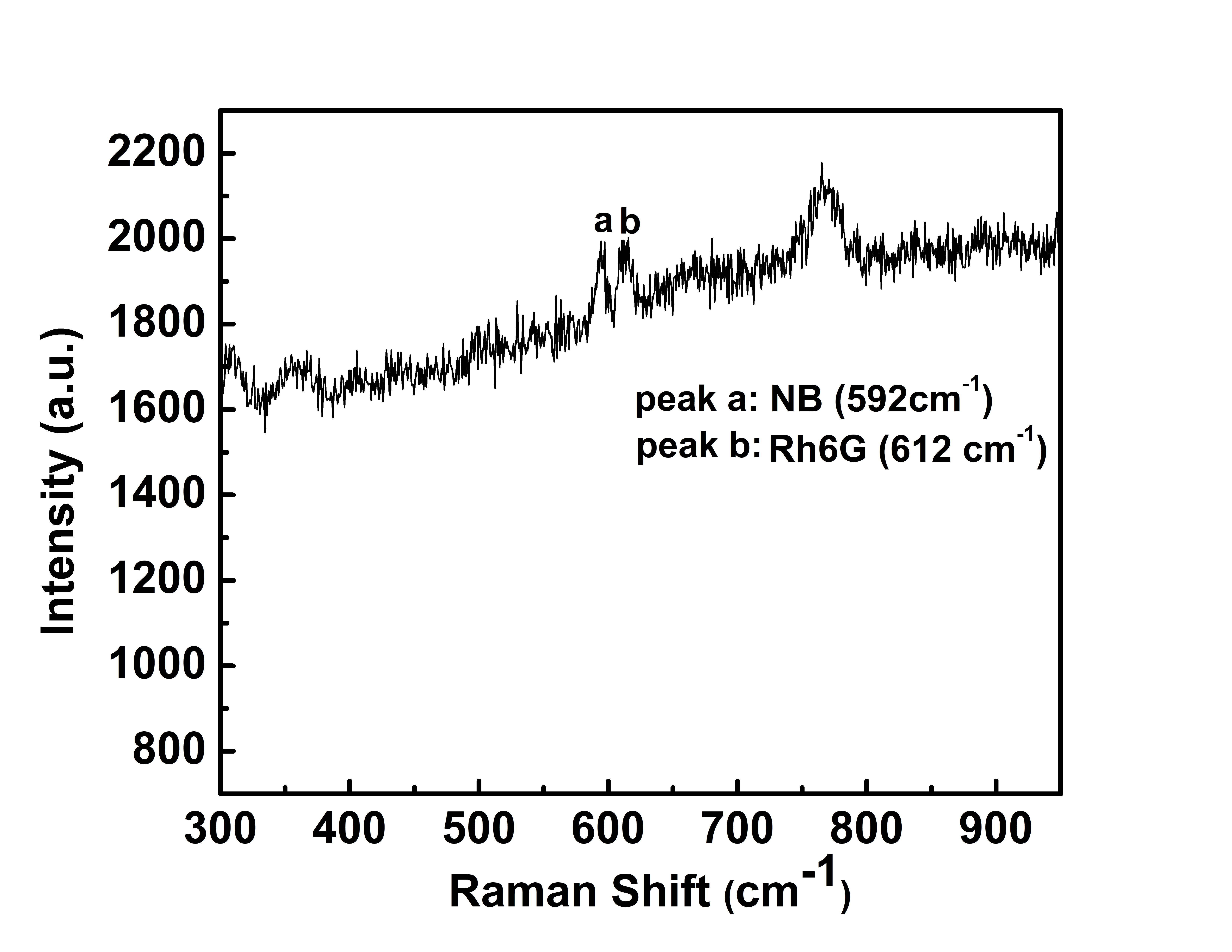}
  \caption{Bi-analyte SERS from assembly of trapped Lee-Meisel AgNPs without adding KCl in the solution. The final concentration of NB and Rh6G molecules in the solution was 5 nM. The exposure time for each spectrum was 5 s. The spectra show two peaks ‘a’ and ‘b’, where ‘a’ is 592 cm$^{-1}$ peak of NB and ‘b’ is 612 cm$^{-1}$ peak of Rh6G. The SERS enhancement was very low as compared to the enhancement obtained from trapped Lee-Meisel AgNPs with added KCl.}
  
\end{figure}

\section{S9: Bi-analyte SERS away from the assembly}

\begin{figure}[H]
\centering
  \includegraphics[width=\linewidth]{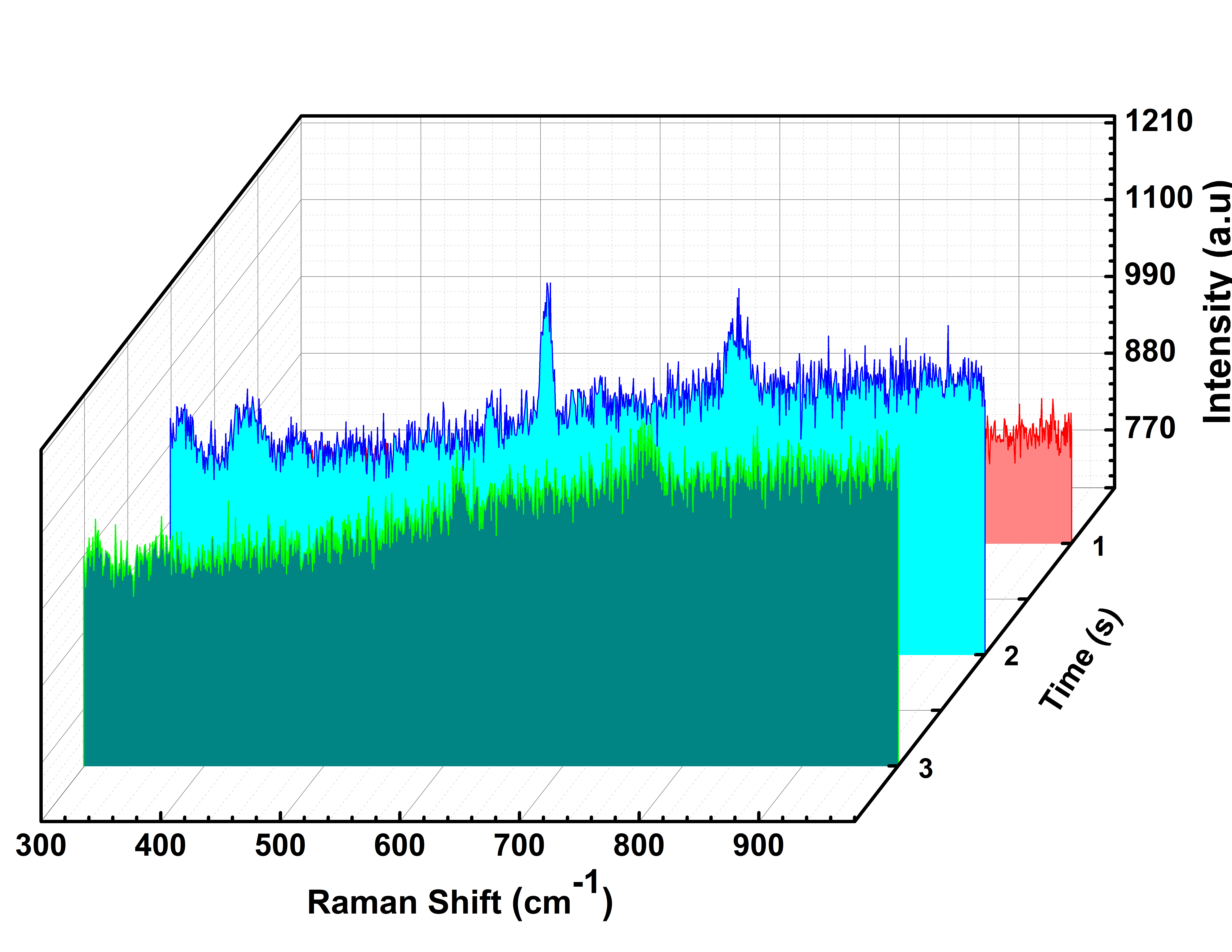}
  \caption{Bi-analyte SERS away from the assembly of Lee-Meisel AgNPs created using a single nanoparticle driven thermoplasmonic trap. The concentration of NB and Rh6G molecules in the final solution was 1 nM. The exposure time for each spectrum was 5 s. The appearance of molecular SERS intensity as a function of time is shown. The peaks of both molecular emission appears less frequently as compared to peaks obtained from the assembly because the number of hotspot away from the assembly was lower as compared to the hotspots present in the assembly. }
  
\end{figure}

\section{S10: Bi-analyte SERS using 532 nm laser}

\begin{figure}[H]
\centering
  \includegraphics[width=\linewidth]{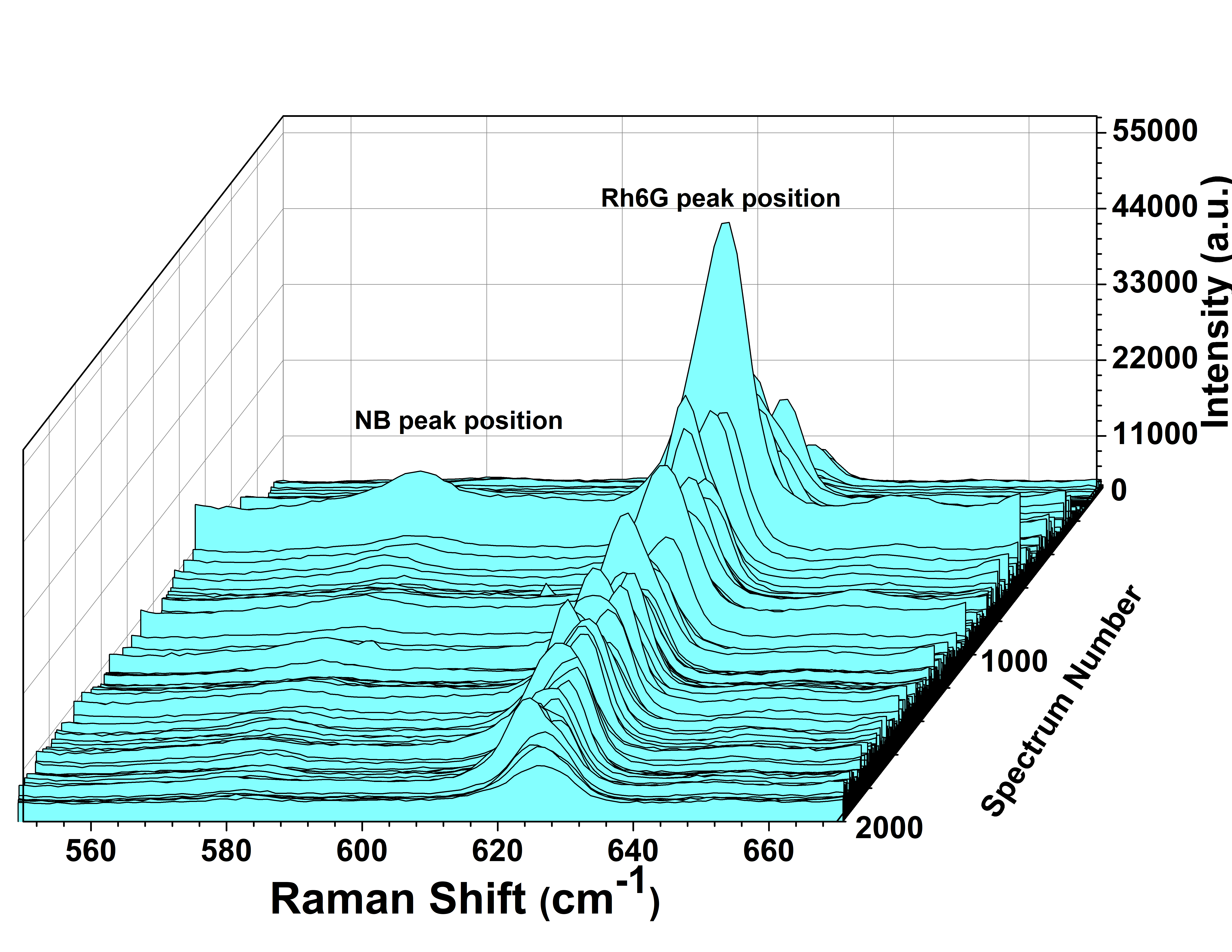}
  \caption{Bi-analyte SERS spectra of bi-analyte solution prepared using 5 nM Rh6G and NB molecules from assembly of Lee-Meisel AgNPs formed using single nanoparticle driven thermoplasmonic trap. The anchor particle was excited with a 532 nm laser for trap and the SERS spectroscopy was performed using the same laser source. The CTAC concentration in the final solution was 10 mM.  The exposure time for each spectrum was 0.3 s and the dwell time between two consecutive spectrums was 1 s. A total of 2000 spectra were taken. The final concentration of KCl salt is 2 mM in the solution and prepared Lee-Meisel AgNPs were diluted by 20 times. The spectra show sharp Rh6G peak but the peak intensity of NB molecules was not good which is generally required for SM-SERS studies.}
  
\end{figure}

\pagebreak

\section{S11: Details on modified principal component analysis}

\begin{itemize}

\item To statistically confirmed the SM-SERS we used MPCA which is developed by P G Etchegoin et al. and can be found in ref\cite{etchegoin2007statistics}.
\item Collected 4000 SERS spectra with 50 wavelengths were arranged in a rectangular matrix (M) of T rows and N columns (4000×50).   
\item After this for each row of the matrix mean intensity was subtracted which gives a zero mean intensity matrix (M’).
\item For the matrix M’ a covariance matrix V of dimension V×V is calculated, from which N Eigenvalues and corresponding N Eigenvectors are calculated and are plotted in figure 4a. In the case of studying SM-SERS using bi-analyte technique, the first two Eigenvalues are principal components.
\item A coefficient matrix C was obtained by matrix operation equivalent to the scalar product of spectra with the first two Eigenvectors.
\item The matrix C was plotted in the coefficient space (shown in figure 4b) and each point in the scatterer plot is an event, with pure events from only one type of molecule is shown along two axes drawn as NB axis and Rh6G axis.
\item To obtain only the positive coefficient that represents the spectra of the molecule a linear transformation was applied to reorient the axis obtained in the scatterer plot in coefficient space.
\item Finally, a histogram is plotted after removing the noisy events from the Figure 4c. The histogram shows the contribution of events from one type of molecules in the total number of spectra. The contribution is plotted in terms of probability of the occurrence of peak of only one type of molecules (as shown in figure 4d).

\end{itemize}

\pagebreak

\section{S12: Details on Supporting Movies}
\begin{enumerate}
  \item Supporting Movie 1: Reversible trapping of 250 nm AuNPs using 150 nm AuNP as anchor trap with 12 mM CTAC solution. The power density used was 0.08mW/$\mu$m$^2$.
  
  \item Supporting Movie 2: No trapping observed in the absence of CTAC with 150 nm AuNP as anchor trap and 250 nm AuNPs in the solution. The power density used was 0.08mW/$\mu$m$^2$.
  
  \item Supporting Movie 3: Reversible trapping of 400 nm AuNPs using 150 nm AuNP as anchor trap with 12 mM CTAC solution. The power density used was 0.08mW/$\mu$m$^2$.
  
  \item Supporting Movie 4: Reversible trapping of Lee-Meisel AgNPs using 150 nm AuNP as anchor trap with 2 mM CTAC solution. The power density used was 0.14mW/$\mu$m$^2$.
  
  \item Supporting Movie 5: Release of Lee-Meisel AgNPs assembly.
  
  \item Supporting Movie 6: Working of tweezer driven by anchor AuNP using 150 nm AuNP as anchor trap and 250 nm AuNPs in the solution with 12 mM CTAC solution. The power density used was 0.08mW/$\mu$m$^2$.
  
  \item Supporting Movie 7: Working of tweezer driven without anchor AuNP with 250 nm AuNPs in the solution and 12 mM CTAC solution. The power density used was 0.08mW/$\mu$m$^2$.
  
  \item Supporting Movie 8: Trapping of single 250 nm AuNP using 150 nm AuNP as anchor trap with 12 mM CTAC solution. The power density used was 0.08mW/$\mu$m$^2$.
  
  \item Supporting Movie 9: Reversible trapping of 250 nm AuNPs using 150 nm AuNP as anchor trap with 15 mM CTAC solution using a 532 nm laser. The power density used was 0.07mW/$\mu$m$^2$.
  
  \item Supporting Movie 10: Unprocessed video of trapping and release of Lee-Meisel AgNPs using 150 nm AuNP as anchor trap with 2 mM CTAC solution. The power density used was 0.14mW/$\mu$m$^2$.
  
\end{enumerate}

\end{document}